# Big Step Greedy Heuristic for Maximum Coverage Problem

Drona Pratap Chandu
Indian Institute of Technology Roorkee
India

## ABSTRACT
This paper proposes a greedy heuristic called Big step greedy heuristic and investigates its application to compute approximate solution for maximum coverage problem. Greedy algorithms construct the solution in multiple steps, the classical greedy algorithm for maximum coverage problem, in each step selects one set that contains the greatest number of uncovered elements. The Big step greedy heuristic, in each step selects $p$ $(1 <= p <= k)$ sets such that the union of selected $p$ sets contains the greatest number of uncovered elements by evaluating all the possible $p$-combinations of given sets. When $p=k$ Big step greedy algorithm behaves like an exact algorithm that computes optimal solution by evaluating all possible $k$-combinations of the given sets. When $p=1$ it behaves like the classical greedy algorithm. The Big step greedy heuristic can be combined with local search methods to compute better approximate solution.

## 1. INTRODUCTION
Maximum coverage problem is to select k sets *{Sx1,Sx2,Sx3,......, Sxk}* from given collection of sets S = *{S1,S2,......, Sn}* such that the number of elements in the union of selected k sets |*Sx1 U Sx2 U* ........... *U Sxk*| is maximum. Maximum coverage problem is a NP-hard problem [1].

Greedy algorithms construct the solution in multiple steps by making a locally optimal decision in each step. The classical greedy algorithm for maximum coverage problem, in each step selects one set that contains the greatest number of uncovered elements. The proposed algorithm called Big step greedy algorithm, in each step selects p (1 <= p <= k) sets such that the union of selected p sets contains the greatest number of uncovered elements by evaluating all possible p-combinations of given sets. Approximation algorithms for Maximum Coverage problem and set covering problem use similar techniques. Grossman and Wool [2] conducted a performance comparison of nine approximation algorithms for set covering problem, and they found that randomized greedy algorithm is the overall best algorithm among the nine approximation algorithms. Results section provides performance comparison of Big step greedy algorithm with randomized greedy algorithm.

## 2. EXISTING APPROXIMATION ALGORITHMS
The classical greedy algorithm for maximum coverage problem is shown in Fig. 1.The classical greedy algorithm starts with empty set cover, and in each step it selects one set that contains the greatest number of remaining elements that are uncovered by current partial solution and adds the selected set to partial solution.

**Algorithm** GMC(*S*,*k*)

*S* : A collection of sets *{S1,S2, ... Sn}*

*k* : Number of sets to be selected from *S*

**begin**

$C \leftarrow \phi$

$W \leftarrow S1 \cup S2 \cup \ldots\ldots \cup Sn$

$S' \leftarrow S$

**while** ($|C| < k$)

Select $T \in S'$ that maximizes $|T \cap W|$

$S' \leftarrow S' \setminus \{T\}$

$C \leftarrow C \cup \{T\}$

$W \leftarrow W \setminus T$

**end while**

return *C*

**end**

**Fig. 1. The classical greedy algorithm for maximum coverage problem**.

The process of adding a set to partial solution is repeated k times to select k sets. Hochbaum and Pathria [3] provides analysis of the classical greedy algorithm for maximum coverage problem.The earlier approximation algorithms [4,5,6] used greedy heuristic for set covering problem. Example 1 explains greedy method with help of a small set collection and the same set collection is used in Example 2 to explain Big step greedy algorithm.

Example 1. Let S = { {a,b,c,d,e,f}, {a,b,c,g,h}, {d,e,f,i,j}, {g,h,i}, {k,l}} be the given collection of sets and K=3.

Assume labels for given sets S1 = {a,b,c,d,e,f}, S2 = {a,b,c,g,h}, S3 = {d,e,f,i,j}, S4 = {g,h,i}, S5 = {k,l}. Initially partial cover C = {}.

In the first step of algorithm, among the five sets S1 has six uncovered elements {a,b,c,d,e,f }and is better than the coverage of sets S2,S3,S4, and S5. So first step selects S1 and now partial cover C = {{a,b,c,d,e,f}}.

In second step, S4 has three uncovered elements {g,h,i}, S2 has two uncovered elements {g,h}, S3 has two uncovered elements {i,j} and S5 has two uncovered elements {k,l} .So second step selects S4 and now partial cover C = {{a,b,c,d,e,f}{g,h,i}}.

In third step, S5 has two uncovered elements {k,l}, S2 has no uncovered elements and S3 has one element{j}. So third step selects S5 and C = { {a,b,c,d,e,f}{g,h,i},{k,l}}

Now |C| = 3 and C covers 11 elements.

Random and probabilistic greedy approximate algorithms [7,8,9] produce better solutions than the classical greedy algorithm for set covering problem. Randomized greedy algorithm used by Grossman and Wool [2] is same as classical greedy algorithm except that ties are broken at random and the basic algorithm is repeated N times and returns the best solution among the N solutions. Computational study by Grossman and Wool [2] shown that randomized greedy algorithm is the best approximation algorithm among the nine algorithms for set covering problem.

Aickelin[10], Beasley and Chu [11] used genetic algorithms for set covering problem. Gomes et al [12] compared four algorithms Round, Dual-LP, Primal-Dual, and Greedy and they concluded that Greedy algorithm performs well among the four algorithms for set covering problem.

Greedy randomized adaptive search procedure (GRASP) [13] is an iterative metaheuristic that can be applied to many combinatorial optimization problems. GRASP in each iteration constructs a feasible solution using randomized greedy adaptive method and it applies local search to find locally optimal solution in the neighbourhood of the constructed solution.

DePuy et al [14] proposed a metaheuristic called Meta-RaPS to solve combinatorial problems and DePuyet al [15] investigated differences between Meta-RaPS and GRASP. Lan et al [16] applied Meta-RaPS for set covering problem and compared with five best algorithms used by Grossman and Wool [2].

Resende [17] applied GRASP for maximum covering problem and shown that GRASP finds near optimal solutions for the majority of the tested problems.

# 3. BIG STEP GREEDY HEURISTIC

Big step greedy heuristic starts with empty set collection, in each step it selects p (1 <= p <= k) sets such that the union of selected p sets contains the greatest number of uncovered elements by evaluating all possible p-combinations of remaining sets and adds the p selected sets to partial set cover.

The process of adding p subsets is repeated k/p times. The last step of the algorithm selects less than p sets when k is not a multiple of p. The Big step greedy algorithm is shown in Fig. 2.

When p=k big step greedy algorithm behaves like an exact algorithm that computes optimal solution by evaluating all possible k-combinations of given sets. When p=1 it behaves like the classical greedy algorithm. When step-size is p the big step greedy algorithm runs in $O((k/p) * |S|^p)$time. Example 2 explains the Big step greedy algorithm with help of the set collection used in Example 1.

**Algorithm** BSGMKC(*S*,*k*,*p*)

*S*: A collection of sets *{S1,S2, … ,Sn}*

*k* : Number of sets to be selected

*p* : step-size of the algorithm

**begin**

*C* ← $\phi$

*W* ← *S1* U *S2* U ……. U *Sn*

**while** (|*C*| < *k*)

**if** ( (*k* - |*C*|) < *p*) **then**

*q* ← *k* - |*C*|

**else**

*q* ← *p*

**end if**

Select *T={T1,T2,....,Tq}*, *T*⊆ *S**C* that maximizes |*W* ∩ *(T1* U *T2* U … .. U *Tq)*|

*W* ← *W* \ (*T1* U *T2* U … .. U *Tq*)

*C* ← *C* U *{T1,T2,...,Tq}*

**end while**

return *C*

**end**

**Fig. 2. Big Step Greedy Algorithm for Maximum Coverage Problem**.

Example 2. Let S = { {a,b,c,d,e,f}, {a,b,c,g,h}, {d,e,f,i,j},{g,h,i }, {k,l}} be the given collection of sets, K=3 and step-size of algorithm is p=2.Assume labels for given sets S1 = {a,b,c,d,e,f}, S2 = {a,b,c,g,h}, S3 = {d,e,f,i,j}, S4 = {g,h,i }, and S5 ={k,l}.As step-size p=2, every step of the algorithm choose two sets such that union of the two selected sets contains the greatest number of uncovered elements.

Initially partial cover C = {}.

In the first step of algorithm, candidates are (S1,S2) , (S1,S3) (S1,S4) (S1,S5) (S2,S3) (S2,S4) (S3,S4)(S3,S5) and (S4,S5), among the ten candidates (S2,S3) is better than all other candidates as S2 U S3 has ten uncovered elements and is greater than that of other candidates. So the first step selects (S2,S3) and now partial cover C = { {a,b,c,g,h} {d,e,f,i,j}}

In second step, it selects only one set instead of two sets because K=3 and two sets S2,S3 are already selected by first step. Candidates are S1, S4, and S5. S5 has two uncovered elements {k,l}, S1 has no uncovered element and S4 has no uncovered elements.

**Algorithm** BestOfBigSteps-1-2-3-4(*S*,*k*)

*S* : A collection of sets *{S1,S2, ... Sn}*

*k* : Number of sets to be selected from *S*

**begin**

*Best* ← $\phi$

**for (*p = 1 to 4*)**

*C* ← BSGMKC*(S,k,p)*

**if** ( | **U** *Best*| < | **U** *C*|) **then**

*Best* ← *C*

**end if**

**end for**

**return** *Best*

**end**

**Fig. 3. Best of Big Steps Algorithm.**

So second step selects S5 and now finally solution C = { {a,b,c,g,h} {d,e,f,i,j} {k,l}} and C covers 12 elements. This is better than the coverage of the sets selected by the classical greedy algorithm in Example 1.

Best of big steps 1,2,3,4 algorithm(BBS-1,2,3,4) shown in Fig. 3 computes four approximate solutions using the big step greedy algorithm with step sizes p=1,2,3,4 and returns the best solution among the four computed solutions.

Big step greedy heuristic does not use local search and big step greedy heuristic can be used in the first phase of GRASP iterations to construct better feasible solutions.

# 4. EXPERIMENTAL RESULTS

Big step greedy heuristic was compared to randomized greedy algorithm, the overall best algorithm among the nine algorithms tested by Grossman and Wool [2].The classical greedy algorithm, the randomized greedy algorithm used by Grossman and Wool[2] and the Big step greedy algorithm for maximum coverage problem were implemented using Java.

Table 1 provides comparison of big step greedy algorithm BS-2 (step size p=2) with classical greedy algorithm and comparison of big step greedy algorithm BS-4 (step size p=4) with classical greedy algorithm.

**Table 1. Greedy Vs Big step greedy on random problem instances**

| \|X\| | Collection Size | Avg Sub set size | k | Number of Problems | Greedy Vs BS-2 | | Greedy Vs BS-4 | |
|---|---|---|---|---|---|---|---|---|
| | | | | | Greedy | BS-2 | Greedy | BS-4 |
| 1000 | 100 | 70 | 10 | 100 | 9 | 31 | 9 | 49 |
| 1000 | 100 | 80 | 10 | 100 | 18 | 30 | 16 | 47 |
| 1000 | 150 | 60 | 15 | 100 | 18 | 36 | 17 | 59 |
| 1000 | 150 | 25 | 5 | 100 | 2 | 11 | 5 | 20 |
| 1000 | 150 | 30 | 5 | 100 | 3 | 3 | 3 | 7 |
| 1000 | 150 | 40 | 5 | 100 | 3 | 11 | 3 | 18 |
| 1000 | 150 | 50 | 5 | 100 | 3 | 12 | 1 | 26 |
| 1000 | 150 | 25 | 5 | 100 | 1 | 1 | 3 | 10 |
| 1000 | 150 | 30 | 5 | 100 | 0 | 9 | 0 | 18 |
| 1000 | 150 | 40 | 5 | 100 | 5 | 5 | 4 | 15 |
| 1000 | 150 | 50 | 5 | 100 | 3 | 19 | 1 | 36 |
| 1000 | 150 | 25 | 10 | 100 | 6 | 10 | 9 | 29 |
| 1000 | 150 | 30 | 10 | 100 | 4 | 19 | 1 | 29 |
| 1000 | 150 | 40 | 10 | 100 | 5 | 19 | 8 | 33 |
| 1000 | 150 | 50 | 10 | 100 | 10 | 27 | 3 | 50 |
| 1000 | 150 | 25 | 15 | 100 | 10 | 21 | 8 | 33 |
| 1000 | 150 | 30 | 15 | 100 | 10 | 26 | 14 | 45 |
| 1000 | 150 | 40 | 15 | 100 | 12 | 31 | 12 | 53 |
| 1000 | 150 | 50 | 15 | 100 | 16 | 27 | 10 | 57 |
| 1000 | 150 | 25 | 20 | 100 | 9 | 29 | 6 | 53 |
| 1000 | 150 | 30 | 20 | 100 | 17 | 21 | 22 | 39 |
| 1000 | 150 | 40 | 20 | 100 | 15 | 35 | 13 | 51 |
| 1000 | 150 | 60 | 5 | 100 | 2 | 14 | 1 | 25 |
| 1000 | 150 | 70 | 5 | 100 | 3 | 12 | 4 | 31 |
| 1000 | 150 | 80 | 5 | 100 | 6 | 12 | 6 | 26 |
| 1000 | 150 | 60 | 10 | 100 | 11 | 29 | 13 | 38 |
| 1000 | 150 | 70 | 10 | 100 | 15 | 30 | 10 | 46 |
| 1000 | 150 | 80 | 10 | 100 | 10 | 26 | 10 | 43 |
| 1000 | 150 | 90 | 5 | 100 | 8 | 23 | 10 | 40 |
| 1000 | 150 | 90 | 10 | 100 | 11 | 43 | 9 | 64 |

In the Table 1, column labeled "|X|" is the number of elements in the universal set, column labeled "Collection Size" is the number of sets in the set collection S of problem instance, column labeled "k" is the number of sets to be

selected from the given collection of sets, column labeled "Number of Problems" is the number of problems used for performance comparison, column labeled "BS-2" under "Greedy Vs BS-2" is the number of problem instances for which Big step greedy heuristic with p=2 is computing better approximate solutions than the classical greedy algorithm and column labeled "Greedy" under "Greedy Vs BS-2" is the number of problem instances for which the classical greedy algorithm is computing better approximate solutions than the Big step greedy heuristic with p=2. The two columns under "Greedy Vs BS-4" have a similar meaning as the columns under "Greedy Vs BS-2".

**Table 2. Randomized Greedy Vs Big step greedy**

| \|X\| | Collection Size | Avg Subset size | k | Number of Problems | R-Greedy Vs BS-3 | | R-Greedy Vs BS-4 | |
|---|---|---|---|---|---|---|---|---|
| | | | | | R-Greedy | BS-3 | R-Greedy | BS-4 |
| 1000 | 100 | 70 | 10 | 100 | 16 | 32 | 19 | 39 |
| 1000 | 100 | 80 | 10 | 100 | 34 | 27 | 30 | 28 |
| 1000 | 150 | 60 | 15 | 100 | 49 | 15 | 37 | 32 |
| 1000 | 150 | 25 | 5 | 100 | 11 | 6 | 11 | 10 |
| 1000 | 150 | 30 | 5 | 100 | 4 | 2 | 10 | 3 |
| 1000 | 150 | 40 | 5 | 100 | 8 | 5 | 8 | 9 |
| 1000 | 150 | 50 | 5 | 100 | 6 | 16 | 4 | 16 |
| 1000 | 150 | 25 | 5 | 100 | 6 | 1 | 9 | 5 |
| 1000 | 150 | 30 | 5 | 100 | 9 | 1 | 4 | 5 |
| 1000 | 150 | 40 | 5 | 100 | 7 | 6 | 8 | 8 |
| 1000 | 150 | 50 | 5 | 100 | 3 | 11 | 1 | 26 |
| 1000 | 150 | 25 | 10 | 100 | 26 | 2 | 26 | 6 |
| 1000 | 150 | 30 | 10 | 100 | 23 | 7 | 17 | 7 |
| 1000 | 150 | 40 | 10 | 100 | 26 | 15 | 35 | 12 |
| 1000 | 150 | 50 | 10 | 100 | 27 | 15 | 22 | 23 |
| 1000 | 150 | 25 | 15 | 100 | 49 | 5 | 41 | 9 |
| 1000 | 150 | 30 | 15 | 100 | 38 | 7 | 36 | 5 |
| 1000 | 150 | 40 | 15 | 100 | 42 | 14 | 36 | 17 |
| 1000 | 150 | 50 | 15 | 100 | 45 | 14 | 36 | 23 |
| 1000 | 150 | 25 | 20 | 100 | 52 | 6 | 39 | 11 |
| 1000 | 150 | 30 | 20 | 100 | 55 | 8 | 49 | 10 |
| 1000 | 150 | 40 | 20 | 100 | 62 | 10 | 51 | 14 |
| 1000 | 150 | 60 | 5 | 100 | 9 | 11 | 6 | 17 |
| 1000 | 150 | 70 | 5 | 100 | 8 | 19 | 12 | 25 |
| 1000 | 150 | 80 | 5 | 100 | 9 | 14 | 7 | 17 |
| 1000 | 150 | 60 | 10 | 100 | 26 | 19 | 22 | 17 |
| 1000 | 150 | 70 | 10 | 100 | 21 | 25 | 20 | 28 |
| 1000 | 150 | 80 | 10 | 100 | 32 | 20 | 26 | 27 |
| 1000 | 150 | 90 | 5 | 100 | 17 | 26 | 15 | 28 |
| 1000 | 150 | 90 | 10 | 100 | 23 | 35 | 24 | 43 |

Between BS-2 (big step with p=2) and the classical greedy algorithm, BS-2 computed better approximate solutions than the classical greedy algorithm for 21% of the problems, and the classical greedy algorithm performed better than BS-2 for 8% of the problems.

Between BS-4 (big step with p=4) and the classical greedy algorithm, BS-4 computed better approximate solutions than the classical greedy algorithm for 36% of the problems, and classical greedy algorithm performed better than the BS-4 for 8% of the problems.

Table 2 provides performance comparison of randomized greedy algorithm with N=20 and big step greedy algorithm BS-3 (with step size p=3) and big step greedy algorithm BS-4 (with step size p=4 )on 3000 randomly generated problem instances.

Between BS-3 (big step with p=3) and randomized greedy algorithm, BS-3 computed better approximate solutions than the randomized greedy algorithm for 13% of the problems, and the randomized greedy algorithm performed better than BS-3 for 25% of the problems.

And between BS-4 (big step with p=4) and randomized greedy algorithm, BS-4 computed better approximate solutions than randomized greedy algorithm for 17% of the problems, and random greedy algorithm performed better than BS-4 for 22% of the problems.

Table 3 provides performance comparison of Best of big steps algorithm(BBS-1,2,3,4) and randomized greedy algorithm on 3000 randomly generated problem instances. BBS-1,2,3,4 algorithm computed better approximate solutions for 22% of the problems and randomized greedy algorithm computed better approximate solutions for 11% of the problems.

**Table 3. Randomized Greedy Vs Best of Big steps 1,2,3,4**

| \|X\| | Collection Size | Avg Subset size | k | Number of Problems | R-Greedy Vs Best Of Big steps | |
|---|---|---|---|---|---|---|
| | | | | | R-Greedy | BBS-1,2,3,4 |
| 1000 | 100 | 70 | 10 | 100 | 7 | 47 |
| 1000 | 100 | 80 | 10 | 100 | 7 | 38 |
| 1000 | 150 | 60 | 15 | 100 | 18 | 35 |
| 1000 | 150 | 25 | 5 | 100 | 2 | 10 |
| 1000 | 150 | 30 | 5 | 100 | 2 | 4 |
| 1000 | 150 | 40 | 5 | 100 | 2 | 10 |
| 1000 | 150 | 50 | 5 | 100 | 0 | 22 |
| 1000 | 150 | 25 | 5 | 100 | 4 | 6 |
| 1000 | 150 | 30 | 5 | 100 | 2 | 5 |
| 1000 | 150 | 40 | 5 | 100 | 3 | 11 |
| 1000 | 150 | 50 | 5 | 100 | 0 | 28 |
| 1000 | 150 | 25 | 10 | 100 | 10 | 7 |
| 1000 | 150 | 30 | 10 | 100 | 10 | 10 |
| 1000 | 150 | 40 | 10 | 100 | 12 | 22 |
| 1000 | 150 | 50 | 10 | 100 | 10 | 28 |
| 1000 | 150 | 25 | 15 | 100 | 32 | 12 |
| 1000 | 150 | 30 | 15 | 100 | 23 | 11 |
| 1000 | 150 | 40 | 15 | 100 | 18 | 23 |
| 1000 | 150 | 50 | 15 | 100 | 25 | 26 |
| 1000 | 150 | 25 | 20 | 100 | 30 | 12 |
| 1000 | 150 | 30 | 20 | 100 | 29 | 16 |
| 1000 | 150 | 40 | 20 | 100 | 36 | 21 |
| 1000 | 150 | 60 | 5 | 100 | 3 | 21 |
| 1000 | 150 | 70 | 5 | 100 | 2 | 31 |
| 1000 | 150 | 80 | 5 | 100 | 2 | 22 |
| 1000 | 150 | 60 | 10 | 100 | 9 | 27 |
| 1000 | 150 | 70 | 10 | 100 | 3 | 36 |
| 1000 | 150 | 80 | 10 | 100 | 11 | 39 |
| 1000 | 150 | 90 | 5 | 100 | 3 | 38 |
| 1000 | 150 | 90 | 10 | 100 | 8 | 55 |

## 5. CONCLUSION

This research proposed a new greedy heuristic called big step greedy heuristic. Big step greedy algorithm was compared with classical greedy algorithm and randomized greedy algorithm[2]. Experiments on many instances of maximum coverage problem shown that big step greedy algorithm with p=2,p=3, and p=4 computes better approximate solutions than the classical greedy algorithm in many cases. As step size p is increased, big step greedy algorithm computed better approximate solutions than the classical greedy algorithm for more percentage of the tested problems.

The randomized greedy algorithm with 20 repetitions computed better approximate solution than the big step greedy algorithm with step size p=3 and with step size p=4 on the average. Best of big steps 1,2,3,4 algorithm computed better approximate solution than the randomized greedy algorithm with 20 repetitions on the average. Best of big steps algorithm proposed in this research can be combined with local search methods to find better approximate solution.